\documentclass[doublecol]{epl2}
\usepackage{epsfig,amsopn}
\usepackage{graphicx}
\usepackage{amsmath,amssymb}
\usepackage{braket}
\usepackage{array}

\newcommand{\eref}[1]{eq.~(\ref{#1})}%
\newcommand{\Eref}[1]{Equation~(\ref{#1})}%
\newcommand{\fref}[1]{fig.~\ref{#1}} %
\newcommand{\Fref}[1]{Figure~\ref{#1}}%

\begin{document}

\title{High Energy Tail of the Velocity Distribution of Driven Inelastic Maxwell Gases}

\author{V. V. Prasad\inst{1}
\and
Sanjib Sabhapandit\inst{1}
\and
Abhishek Dhar\inst{2}
}

\shortauthor{V. V. Prasad,
Sanjib Sabhapandit,
and
Abhishek Dhar 
}

\institute{ 
\inst{1} Raman Research Institute, Bangalore Bangalore
  560080, India \\
\inst{2} International Centre for
   Theoretical Sciences, TIFR, Bangalore 560012, India
}


\abstract{ A model of homogeneously driven dissipative system,
  consisting of a collection of $N$ particles that are characterized
  by only their velocities, is considered.  Adopting a discrete time
  dynamics, at each time step, a pair of velocities is randomly
  selected. They undergo inelastic collision with probability
  $p$. With probability $(1-p)$, energy of the system is changed by
  changing the velocities of both the particles independently
  according to $v\rightarrow -r_w v +\eta$, where $\eta$ is a Gaussian
  noise drawn independently for each particle as well as at each time
  steps. For the case $r_w=- 1$, although the energy of the system
  seems to saturate (indicating a steady state) after time steps of
  $O(N)$, it grows linearly with time after time steps of $O(N^2)$,
  indicating the absence of a eventual steady state. For $ -1 <r_w
  \leq 1$, the system reaches a steady state, where the average energy
  per particle and the correlation of velocities are obtained
  exactly. In the thermodynamic limit of large $N$, an exact equation
  is obtained for the moment generating function. In the limit of
  nearly elastic collisions and weak energy injection, the velocity
  distribution is shown to be a Gaussian.  Otherwise, for $|r_w| < 1$,
  the high-energy tail of the velocity distribution is Gaussian, with
  a different variance, while for $r_w=+1$ the velocity distribution
  has an exponential tail.  }

\maketitle

As we observe in everyday life, for example when we drop a marble ball on the
floor or play billiard, granular matter dissipates energy through
inelastic collisions.  A box of marbles needs to be shaken
continuously in order to observe movement of the constituents in the
system. If the supply of energy from the outside is stopped, such
systems comes to the rest in a short time. When the input of energy to
the system (``heating'') compensates the energy loss due to collisions
(with walls as well as with other particles), one expects an inelastic
``gas'' to reach a nonequilibrium steady-state.  It is tempting to
define an ``effective temperature'' by the average kinetic energy per
particle.  A natural question is whether the steady-state velocity
distribution is Maxwell-Boltzmann or something else.

In this letter we consider a much studied model of driven inelastic
granular gases, namely the inelastic Maxwell model. We first point out
a subtle effect that has been overlooked in earlier studies --- that
collisions with walls are necessary for attaining a steady state.  We
obtain an exact evolution equation for the velocity covariance
[\eref{x-recursion}] which has an explicit solution. We also obtain
several other results in the thermodynamic limit, including an exact
equation for the moment generating function [\eref{recursion}], exact
recursion relations for all moments [\eref{moment ratio}] and
determination of the high-energy tails of the velocity distribution
[\eref{LD tail}].

The steady state velocity distributions measured in experiments show
deviation from the Maxwellian statistics as well as approach to
Gaussian distribution, depending on the experimental
conditions~\cite{Olafsen:99, Losert:99, Kudrolli:00, Rouyer:00,
  Aranson:02,Kohlstedt:05}.  Based on the kinetic theory, an analysis
by van Noije and Ernst~\cite{Noije:98} for a uniformly heated
inelastic hard sphere gas, predicts a non-Maxwellian tail $P(v)\sim
\exp(-A |v|^\alpha)$ with $\alpha=3/2$, of the velocity
distribution. Barrat \textit{et al.}~\cite{Barrat:02} find $\alpha=3$
in the limit of vanishing inelasticity (see also~\cite{Benedetto:98}).
On the other hand, the numerical study of a two-dimensional driven
inelastic gas~\cite{vanZon:04} shows a wide range of $\alpha$
($\alpha<2$) depending on the various system parameters, instead of a
universal value.  A one-dimensional model of granular system
consisting of Brownian particles subject to inelastic mutual
collisions, shows a crossover from Gaussian to non-Gaussian
distribution, as the ratio of the Brownian relaxation time to the mean
collision time is increased~\cite{puglisi:98}. A model of a
  driven dissipative system was studied in \cite{srebro:04} where the
  energy moments could be computed exactly.

\begin{figure}
\centerline{\includegraphics[width=\hsize]{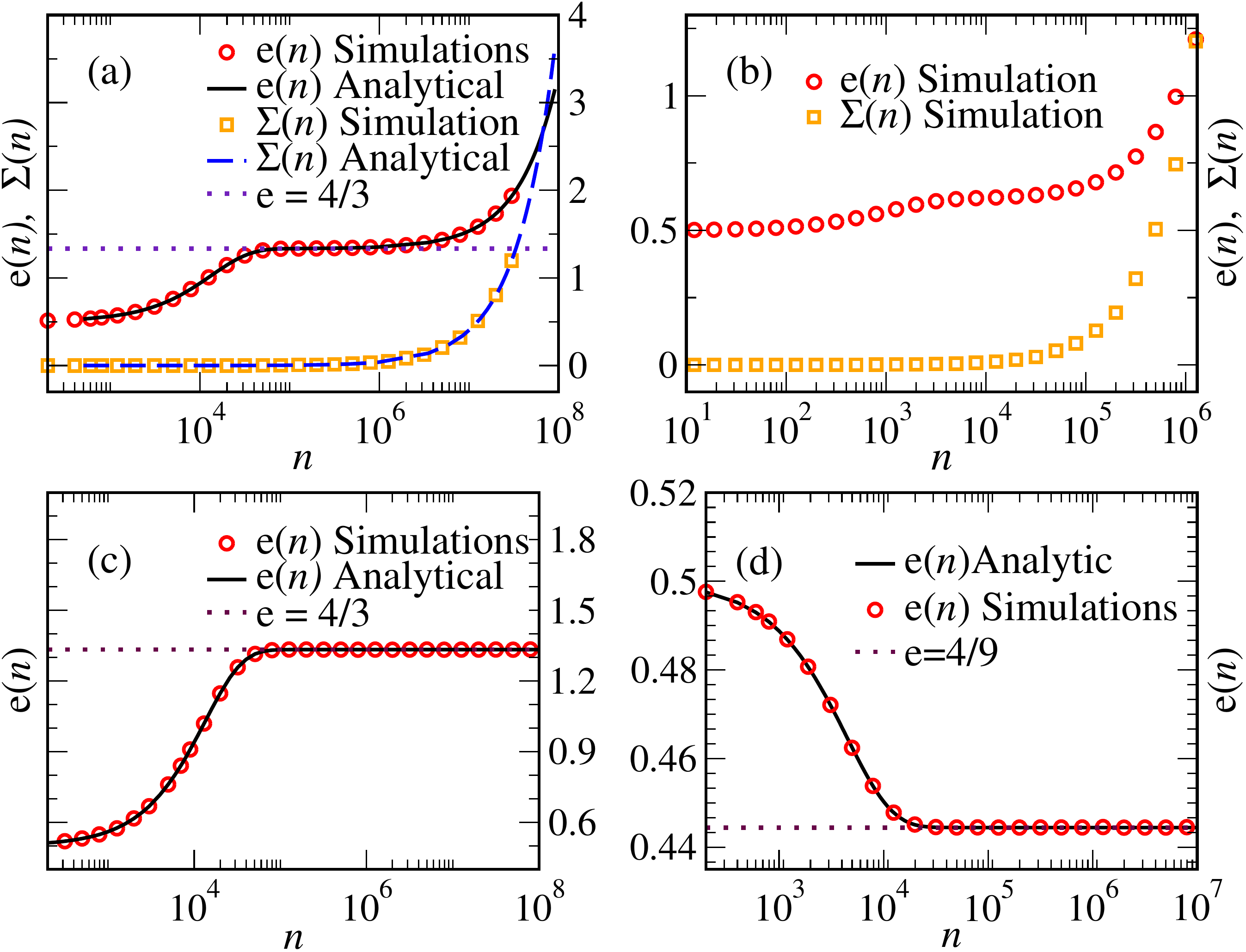}}
\caption{(Color Online) (a) Simulation results for the evolution of
  the energy per particle $\mathrm{e}(n)$ (red circles) and the
  velocity correlation $\Sigma(n)$ (orange squares) of $N=5000$
  particles system driven by adding uncorrelated white noise
  ($r_w=-1$), for $p=1/2$ and $\epsilon=1/4$.  Analytical results are
  shown by (black) solid and (blue) dashed lines respectively. The
  $pseudo$ steady state value is shown by the (magenta) dotted
  line. (b) Simulation results for $\mathrm{e}(n)$ (red circles) and
  $\Sigma(n)$ (orange squares) with $N=1000$, $p=1/2$ and
  $\epsilon=1/4$, for $\delta=1$ case. (c) Simulation (red circles) as
  well as analytical (black solid line) results for $\mathrm{e}(n)$
  for $N=5000$, with the dynamics given by \eref{eqn wall dissipation
    different rates}, with $p=1/2$, $\epsilon=1/4$, and $r_w=+1$. The
  steady state value is shown by the (magenta) dotted line. (d)
  Simulation (red circles) as well as analytical (black solid line)
  results for $\mathrm{e}(n)$ for $N=5000$, with the dynamics given by
  \eref{eqn wall dissipation different rates}, with $p=1/2$,
  $\epsilon=1/4$, and $r_w=1/2$. The steady-state value is shown by
  the (magenta) dotted line.}
\label{Fig1}
\end{figure}

Kinetic theory approach to homogeneous granular gases involves setting
up the Boltzmann equation for the single-particle velocity
distribution, under the \emph{molecular chaos} hypothesis, which
replaces the two-particles velocity distribution function by the
product of two single-particle probability density functions (PDFs)
$P(v_1,v_2)=P(v_1)P(v_2)$.  For
hard spheres, the collision rate is proportional to $|v_1-v_2|^\delta$
with $\delta=1$. In the absence of external driving it has been shown  that the velocity distribution has an exponential tail $P(v)\sim \exp ({-|v|/v_0(t)})$ where $v_0(t)$ is the time-dependent thermal speed \cite{episov97}.   
In the driven case, uniform heating is modeled by adding a
diffusive term $D\partial_v^2 P(v)$ to the Boltzmann equation.  Ben-Naim and Krapivsky introduced a simpler model,
called inelastic Maxwell model~\cite{Ben-naim:00}, where the collision
rate is independent of the velocities of the colliding particles,
i.e., $\delta=0$.  In this model, an exponential tail $P(v)\sim
\exp(-A|v|)$ for the velocity distribution has been
obtained~\cite{Santos:03, Antal:02, Marconi:02}, while in
  the limit of vanishing inelasticity, the distribution becomes
  Gaussian \cite{Barrat:07}.

The inelastic Maxwell model is perhaps the simplest model of granular
gases.  The modeling of velocity change due to an external forcing
(heating) by the uncorrelated white noise, $dv_j/dt=\eta_j(t)$, (which
corresponds to the diffusion term in the Boltzmann equation) injects
energy, on average, to the system.  However, as it turns out, this
energy input cannot be balanced by the energy dissipation due to
inelastic collisions, and eventually, the average energy of the system
increases linearly with time and \emph{hence there is no steady state}
[see \fref{Fig1}(a)]. This is true even for $\delta > 0$, as seen
from \fref{Fig1}(b).  This is because the inter-particle collision
conserves total momentum while the driving noise causes the total
momentum to perform a random walk --- hence the average energy of the
system grows linearly with time.  One usually justifies this model of
heating by moving to the center of mass velocity frame~\cite{Williams:96,
Ben-naim:00}.  However, it is not the experimental situation and hence
is unsatisfactory. In fact, as we show below, the external driving is better modeled by a Ornstein-Uhlenbeck process, if
one assumes that the external forcing is due to collisions
with a wall with a random velocity.  While, similar forcing mechanism
has been considered earlier for inelastic gases~\cite{puglisi:98,
puglisi:99, Marconi:02, Costantini:07}, the form of the high-energy
tail has not been obtained.  In this letter, we obtain several exact
results, including the high-energy tail of the velocity distribution.

Here for simplicity, we consider the dynamics in discrete time
steps. It is straightforward to obtain the continuum version by taking
appropriate limits \cite{Prasad}.  Following Ref.~\cite{Ben-naim:00},
ignoring the spatial structure, we consider a collection of $N$
identical particles that are characterized by their velocities $v_i$,
where $i=1,2, \dotsc, N$. The initial velocities are chosen
independently from a Gaussian distribution. At each time step, two
particles are chosen at random, and with probability $p$ they undergo
inelastic collision and with probability $1-p$ they are subjected to
independent external forcing.  The two-body inelastic collisions
change the velocities of the selected pair from $(v_i,v_j)$ to $(v_i',
v_j')$ according to
\begin{equation}
\begin{split}
v'_{i} = \epsilon v_i + (1 - \epsilon) v_{j},\\
v'_{j} = (1 - \epsilon) v_{i} + \epsilon v_j.
\end{split}
\end{equation} 
Here $\epsilon = (1-r)/2$,  with $r$ being the coefficient of
restitution defined by $(v'_i-v'_j)=-r(v_i-v_j)$. The elastic
collision corresponds to $r=1$. Although $r=0$ corresponds to the
completely inelastic case, in this model the particles do not stick to
each other and merely possess the same velocity after collision.
Therefore, collision conserves both the number of particles and total
momentum ($v_i'+v_j'=v_i+v_j$). In fact, as a model of a dissipative
system, we can take $\epsilon\in(0,1)$.

\setcounter{equation}{5}
\begin{widetext}
\begin{equation}
\begin{split}
&X_n= \left[\begin{array}{>{\displaystyle}c}
\mathrm{e}(n) \\ \Sigma(n)
\end{array} 
\right], ~\qquad
C= \left[\begin{array}{>{\displaystyle}c}
(1-p)\frac{\sigma^{2}}{N} \\ 0
\end{array} 
\right],\quad\text{and}\\[3mm]
&R=\left[\begin{array}{>{\displaystyle}c>{\displaystyle}c} 
1- \frac{\left[4p\epsilon(1-\epsilon)+2(1-p)(1-r^{2}_{w})\right]}{N} & \frac{2p\epsilon(1-\epsilon)}{N}~\\ \\
\frac{8p\epsilon(1-\epsilon)}{N(N-1)} & 1-
\frac{\left[4p\epsilon(1-\epsilon)-2(1-p)(1+r_{w})^{2} +
    4(N-1)(1-p)(1+r_{w})\right]}{N(N-1)}\end{array}\right]
\label{wide1}
\end{split}
\end{equation}
\begin{align}
\label{wide2a}
{\rm
e}&=\frac{(\sigma^2/2)\left[2\epsilon(1-\epsilon)+\gamma(1-r_w^2)+2(N-2)\gamma(1+r_w)\right]}{4\epsilon(1-\epsilon)(1-r_w^2)+\gamma(1-r_w^2)^2+(N-2)(1+r_w)[4\epsilon(1-\epsilon)+2\gamma(1-r_w^2)]} ~, \\
\label{wide2b}
\Sigma &=\frac{2\sigma^2
\epsilon(1-\epsilon)}{4\epsilon(1-\epsilon)(1-r_w^2)+\gamma(1-r_w^2)^2+(N-2)(1+r_w)[4\epsilon(1-\epsilon)+2\gamma(1-r_w^2)]}~.
\end{align}
\end{widetext}
\setcounter{equation}{1}

As in Ref.~\cite{Ben-naim:00}, we first apply the external forcing by
adding uncorrelated white noises to the velocities ($v'=v+\eta$) of
the selected pair independently.  In this case, starting from an
initial state of uncorrelated velocities, we find that~\cite{Prasad}
for large $N$, after time steps of $O(N)$ the system reaches a
steady-state-like state where the average energy per particle
saturates to a fixed value [\fref{Fig1}(a)]. Although, correlation
between velocities builds up over time through collisions, it is
$O(1/N)$ in this \emph{pseudo} steady state and therefore remains
negligible. Consequently, the molecular chaos hypothesis,
$P(v_i,v_j)=P(v_i)P(v_j)$, is justified in the \emph{pseudo} steady
state.  However, eventually after time steps of $O(N^2)$, the
correlation between velocities no longer remains negligible --- both
the correlation and the average energy per particle grow linearly with
time [\fref{Fig1}(a)]. Therefore, the system no longer has
a \emph{true} steady state.

If one assumes that the external forcing is due to  collision
of a particle with a vibrating wall, then the post-collision velocity
$v'$ of a particle is related to its pre-collision velocity $v$ and
the velocity of the wall $V_w$ as
\begin{math}
v'=-r_w v + (1+r_w) V_w,
\end{math}
where $r_w$ is the coefficient of restitution for the particle-wall
collision and the velocity of the wall is assumed to be unchanged in
the collision.  Since, the velocity of the wall is random, we denote
the term $(1+r_w)V_w$ by a random variable $\eta$ to get
\begin{math}
v'=-r_w v + \eta.
\end{math}
In an appropriate limit this becomes equivalent to an
Ornstein-Uhlenbeck process in continuous time.  We will see that this
leads to a steady state for $-1<r_w\leq1$.

The dynamics $(v_i,v_j)\rightarrow (v'_i,v'_i)$ at each time step, for
the pair selected at random, is given by
\begin{equation}
\begin{split}
\label{eqn wall dissipation different rates}
v^{\prime}_{i} &= \chi \bigl[\epsilon v_i + (1 - \epsilon) v_{j}\bigr]
+ (1-\chi)\bigl[-r_{w} v_{i} + \eta_{i}\bigr],\\
v^{\prime}_{j} &= \chi \bigl[\epsilon v_j + (1 - \epsilon) v_{i}\bigr]
+ (1-\chi)\bigl[-r_{w} v_{j} + \eta_{j}\bigr],
\end{split}
\end{equation}
where $\chi$ is a random number which takes values either $1$ or $0$
with probability $p$ and $1-p$ respectively, independently at each
step. The noises $\{\eta\}$ at different time steps are drawn independently 
from a Gaussian distribution with zero mean and variance
$\langle \eta_i\eta_j\rangle =\sigma^2\delta_{ij}$.

Let $v_{i}(n)$ be the velocity of the $i-{\rm th}$ particle at the
$n-{\rm th}$ time step.  The average energy per particle of the system
and the velocity correlation between any two particles, at the $n-{\rm
  th}$ time step are defined as
\begin{align}
{\rm e}(n) &= \frac{1}{2N}\; \sum^{N}_{i=1}\; \langle v^2_{i}(n)\rangle~, 
\\
\text{and}~~\Sigma(n) &= \frac{1}{N(N-1)}\; \sum_{i\neq j}\; \langle v_{i}(n) v_{j}(n)\rangle~,
\end{align}
respectively. It turns out that they  satisfy the following exact recursion relation:
\begin{equation}
X_{n} = RX_{n-1} + C,
\label{x-recursion}
\end{equation}
where $X_n$, $R$, and $C$ are given by 
$$ see~eq.(\ref{wide1}) ~above$$

\setcounter{equation}{8}

\Eref{x-recursion} has the exact solution $X_{n} = R^{n} X_{0} +
\sum^{n-1}_{l=0}\; R^{l} C$, with the initial conditions
$X_{0}=[\mathrm{e}_0, 0]^T$. The two eigenvalues of $R$ can be
evaluated explicitly. For $r_w=-1$, the eigenvalues are $1$ and
$1-p(1-r^2)/(N-1)$. The unit eigenvalue leads to an eventual linear
increase of $X_n$.  For $r_w \neq -1$, both eigenvalues have absolute
values less than unity, and hence in the $n\rightarrow \infty$ limit,
$X_n\rightarrow X_\infty$, independent of $n$, as seen in
\fref{Fig1}(c) and \fref{Fig1}(d). Thus, the system reaches a steady
state. The energy and correlations in the steady state are given
exactly by 
$$ see~~eqs.~\eqref{wide2a}~and~\eqref{wide2b} ~above$$
respectively, where $\gamma=(1-p)/p$, is the ratio of the injection to
the collision rate.
 In the large $N$ limit we get
\begin{equation}
\mathrm{e}=\frac{
\gamma\sigma^2}{4\epsilon(1-\epsilon)+2\gamma(1-r_w^2)}+O(N^{-1})
\label{energy}
\end{equation}
 and $\Sigma= O(N^{-1})$. Therefore, in the steady state, in the
 thermodynamic limit $N \to \infty$, the correlations between the
 velocities vanish. In this limit, it is easy to show that, the moment
 generating function $Z(\lambda)=\langle \exp(-\lambda v)\rangle$ of
 the steady-state velocities satisfies the equation
\begin{equation}
Z(\lambda)=p Z(\epsilon\lambda)Z([1-\epsilon]\lambda) 
+(1-p) Z(r_w \lambda)f(\lambda),
\label{recursion}
\end{equation}
where $f(\lambda)=\exp(\lambda^2\sigma^2/2)$ and we have used the fact
that $Z(-\lambda)=Z(\lambda)$ for even distribution. Note that
$Z(\lambda)=1+\mathrm{e} \lambda^2 +\dotsb$ as $\lambda\rightarrow 0$.

\begin{figure}
\centerline{\includegraphics[width=.8\hsize]{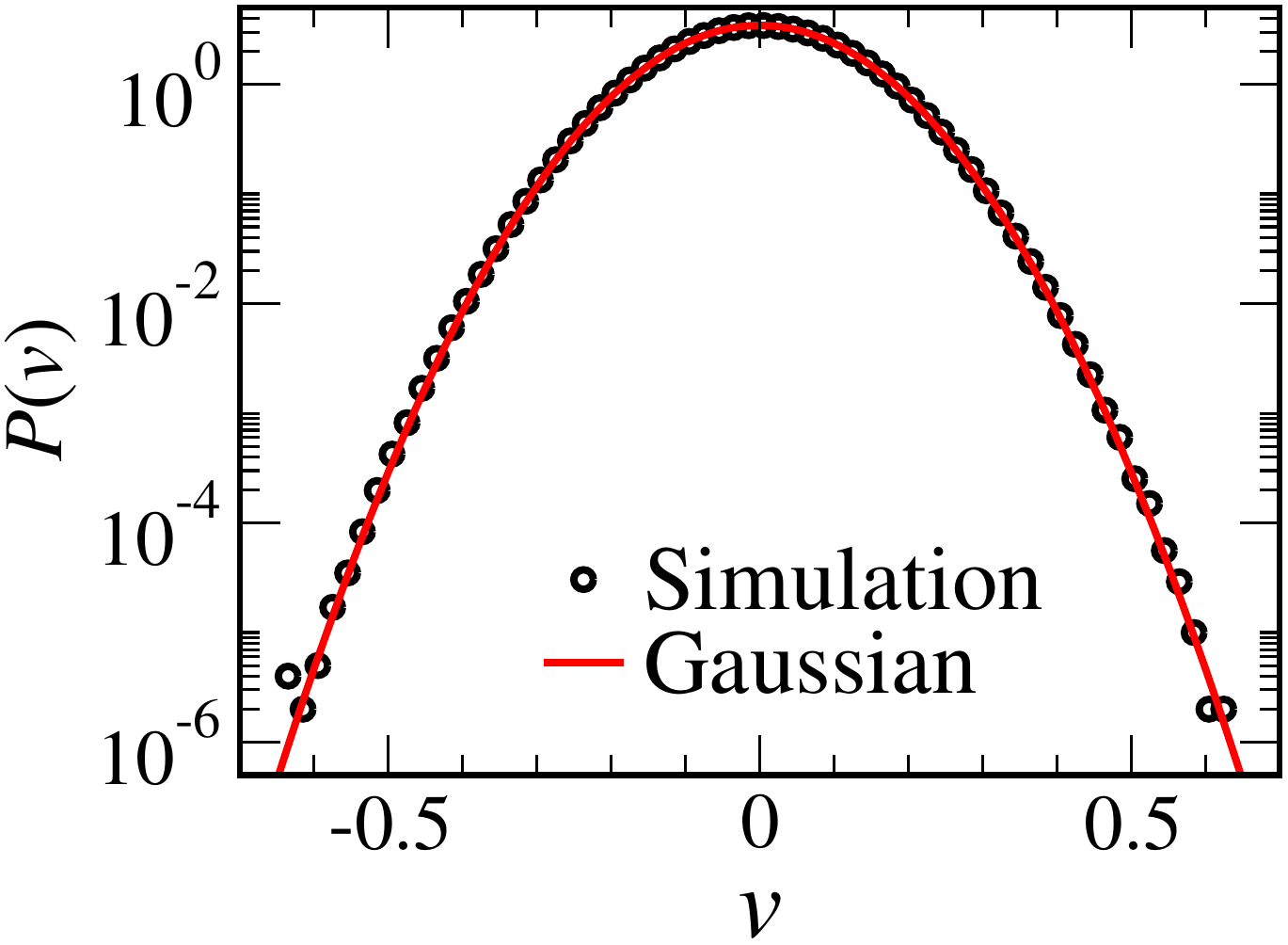}}
\caption{\label{Fig2}
Simulation (black
  circles) and analytical (red solid line) results for steady state
  velocity distribution in the near-elastic and weak energy injection
  limit for $N=100$ with $p=0.5$, $\epsilon=0.005$, $r_w=0.99$, and
  $\sigma=0.02$.}
\end{figure}

In the near-elastic and weak energy injection limit:
$\epsilon\rightarrow 0$, $r_w=(1-\theta)\rightarrow 1$,
$\sigma\rightarrow 0$, while keeping $\sigma^2/\epsilon$ and
$\theta/\epsilon$ fixed, using the Taylor expansion in
\eref{recursion} we get
\begin{equation}
\frac{dZ}{d\lambda}=\lambda\Delta^2 Z(\lambda), ~~\text{where}~
\Delta^2=
\frac{\gamma\sigma^2/\epsilon}{2\bigl[1+\gamma \theta/\epsilon\bigr]}.
\label{elasticZ}
\end{equation}
Note from \eref{energy} that $\langle v^2\rangle \rightarrow \Delta^2$ in this
limit.  Evidently, the solution of \eref{elasticZ} is
$Z(\lambda)=\exp(\lambda^2\Delta^2/2)$, which implies the Gaussian
velocity distribution
\begin{equation}
P(v)=\frac{1}{\sqrt{2\pi\Delta^2}}\exp\left(-\frac{v^2}{2\Delta^2}\right).
\end{equation}
\Fref{Fig2} shows a comparison of this result with numerical simulation.

\begin{figure}
\centerline{\includegraphics[width=.8\hsize]{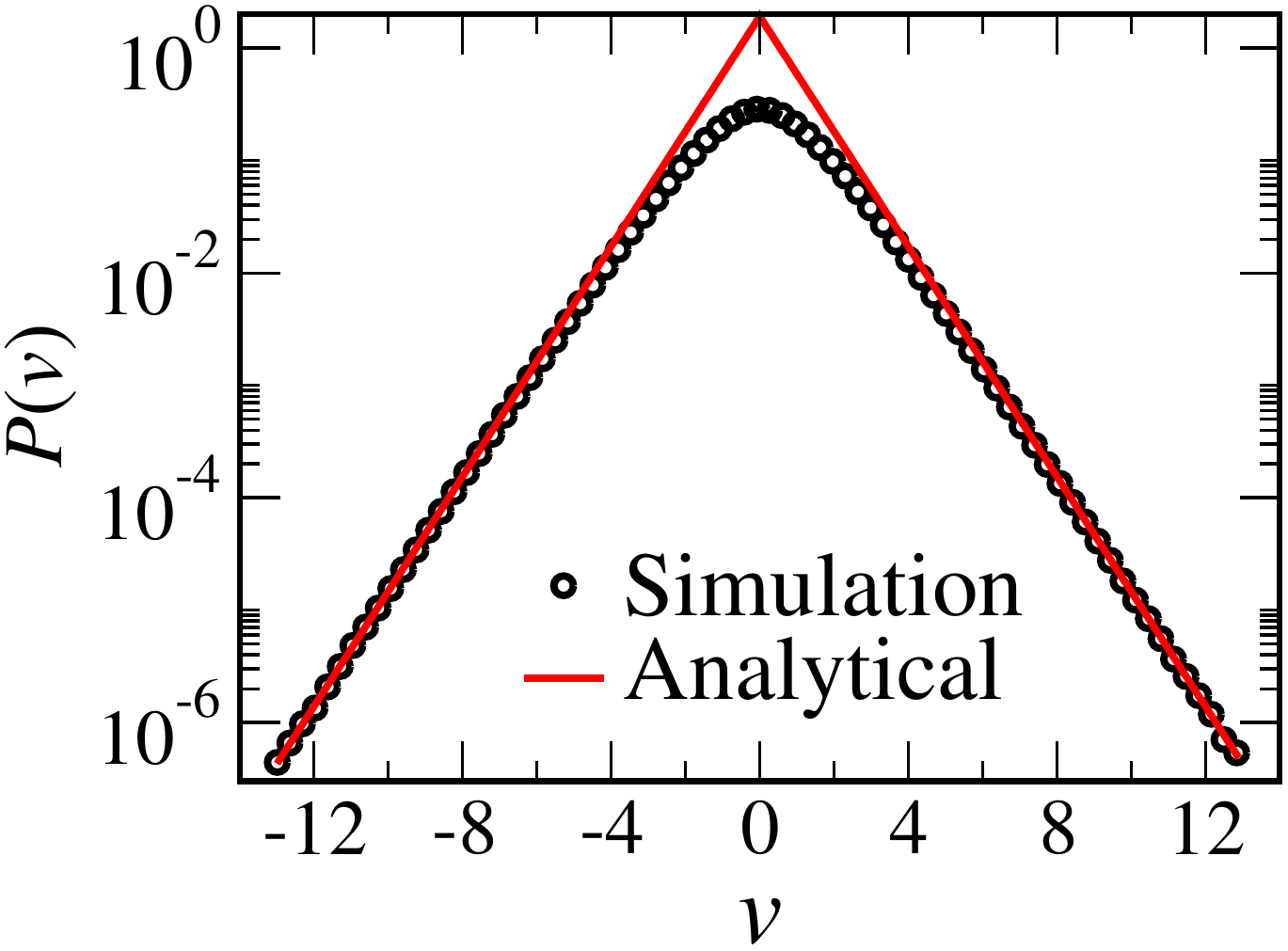}}
\caption{\label{Fig3} Simulation (black circles) and analytical (red
  solid line) results for steady state velocity distribution for
  $r_w=+1$.  The other parameters are $N=100$, $p=0.5$,
  $\epsilon=1/4$, and $\sigma=1$.}
\end{figure}

In general, we are not able to find the exact solution of
\eref{recursion}.  For the case $r_w=+1$, we have
\begin{equation}
Z(\lambda)=\bigl[1-(1-p)~f(\lambda)\bigr]^{-1}\, p\, Z(\epsilon \lambda) Z([1-\epsilon]\lambda)~,\label{rweq1}
\end{equation}
which can be solved by iteration \cite{Prasad}. The tail of the
velocity distribution is determined by the pole closest to the origin
$\lambda_0=\pm \sqrt{-2 \ln (1-p)}/\sigma$, which comes from the
factor $[1-(1-p)~f(\lambda)]^{-1}$ in \eref{rweq1}. This gives rise to
exponential tails $P(v) \sim A(\epsilon) \exp (- |\lambda_0| |v|)$,
where the prefactor $A(\epsilon)$ is known explicitly. A comparison
with the numerical simulation gives very good agreement as shown in
\fref{Fig3}. We note that \eref{recursion} is symmetric for $r_w
\leftrightarrow -r_w$. However this equation has been derived under
the assumption of a steady state. Therefore for the case $r_w=-1$, it
is valid \emph{only} in the pseudo-steady state in which case one
indeed finds an exponential tail \cite{Ben-naim:00,Santos:03,Antal:02,
  Marconi:02}.

\begin{figure*}
\includegraphics[width=\hsize]{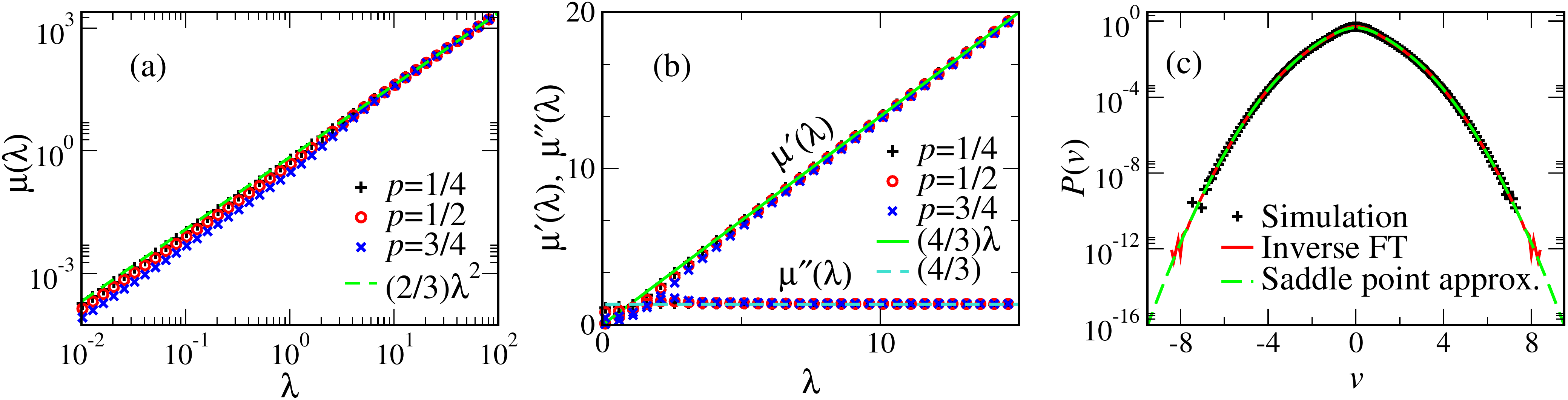}
\caption{(Colors Online) (a) $\mu(\lambda)$ for  $\epsilon=1/2$, and
$r_w=1/2$ for three different $p$. These curves asymptotically
approach the function $(2/3)\lambda^2$, shown by the (green) dashed
line. (b) The first and second derivatives of $\mu(\lambda)$ and their
asymptotic values (green solid line and cyan dashed line
respectively). The same symbols and colors are used for the same $p$
values for both $\mu'(\lambda)$ and $\mu''(\lambda)$. (c) The velocity
distribution for the dynamics \eref{eqn wall dissipation different
rates} for $p=1/2$, $\epsilon=1/2$ and $r_w=1/2$ from simulation
(black ``+") compared with results calculated from the exact inverse
Fourier transform of $Z_c(k)$ (red solid line) as well as from the
saddle point approximation (green dashed line), given
by \eref{SP-soln}. }
\label{Fig4}
\end{figure*}

For $|r_w|<1$ it is difficult to obtain the tails accurately from
direct numerical simulations, as it would require large number of
realizations. \Eref{recursion} can be numerically solved and then
inverted numerically to obtain $P(v)$. It is convenient to use the
characteristic function $Z_c(k)=Z(ik)$.  For the special case
$\epsilon=1/2$ and $r_w=1/2$, \eref{recursion} has a simpler form
\begin{equation}
Z_c(k)=p Z_c^2(k/2)+ (1-p)Z_c(k/2)\exp(-k^2\sigma^2/2).
\end{equation}
This is useful, as to compute $Z_c$ for any $k$ one only requires its
value at $k/2$. This ``linear structure'', as opposed to the ``tree''
in the general case, is useful to efficiently compute $Z_c(k)$
numerically while using the initial condition $Z_c(k)=1-\mathrm{e}k^2$
for $k \ll 1$.  Finally, numerically computing the inverse Fourier
transform of $Z_c(k)$ gives the velocity distribution.  This is
compared with simulation results in \fref{Fig4}(c). From the numerical
evaluation, we observe the tail $P(v)\sim \exp(-A \mid
v\mid^{\alpha})$, with $\alpha$ gradually increasing (but $<2$) as we
go towards higher and higher the velocities.

For $\alpha >1$, the function $Z(\lambda)$ is analytic.  If
$Z(\lambda)$ is known, then the large deviation tail of the velocity
distribution can be obtained by the saddle point approximation
\begin{equation}
P(v)\approx\frac{\exp\bigl[\mu(\lambda^*) +\lambda^*
v\bigr]}{\sqrt{2\pi|\mu''(\lambda^*)|}}~,
\end{equation}
where $\mu(\lambda)=\ln
Z(\lambda)$ and the saddle point $\lambda^*(v)$ is implicitly given by
the equation $\mu'(\lambda^*)=-v$. 
A careful saddle-point analysis \cite{Prasad} shows that
the first term on the right
hand side of \eref{recursion} becomes negligible compared to the left
hand side, for $\lambda$ near $\lambda^*(v)$ for large $v$. The remaining terms in \eref{recursion} gives the Gaussian distribution 
\begin{equation}
P(v)\approx \sqrt{\frac{1-r_w^2}{2\pi \sigma^2}}
\,\exp\left[-\frac{v^2}{2\sigma^2}
(1-r_w^2)\right].
\label{LD tail}
\end{equation} 
Therefore, the high-energy tail is governed, not by the inelastic
collisions amongst the particles, but by the collisions of the particles
with the ``wall''.

To verify the above result, we numerically compute $Z(\lambda)$
from \eref{recursion} for $\epsilon=1/2$, $r_w=1/2$, and $\sigma=1$,
for which $b=2/3$. It is clear from \fref{Fig4}(a) that
$\mu(\lambda)\sim b\lambda^2$ for large $\lambda$. From
$\mu(\lambda)$, we also numerically compute $\mu'(\lambda)$ and
$\mu''(\lambda)$ [\fref{Fig4}(b)]. Now, each value of $\lambda$ corresponds to a
velocity $v=-\mu'(\lambda)$, whose PDF, under the saddle point
approximation, is numerically obtained using
\begin{equation}
P(v=-\mu'(\lambda)) \approx \frac{1}{\sqrt{2\pi \mu''(\lambda)}}\exp
\bigl[\mu(\lambda) -\lambda \mu'(\lambda)\bigr].
\label{SP-soln}
\end{equation}
\Fref{Fig4}(c) compares this with  the simulation result as well as
with the distribution obtained from the exact numerical inverse
Fourier transform of $Z_c(k)$.

\begin{figure}
\centerline{\includegraphics[width=.8\hsize]{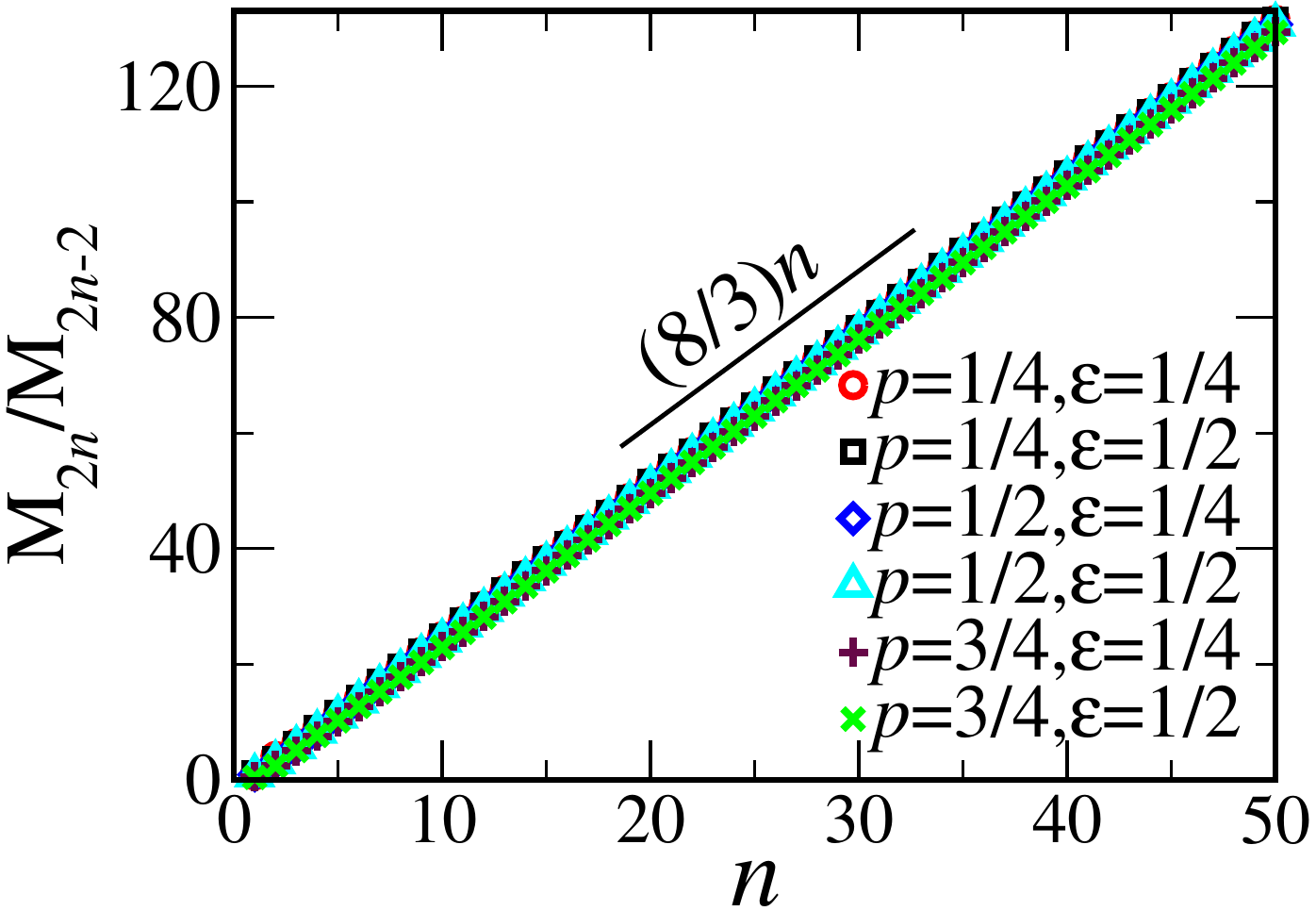}}
\caption{The ratios of successive even moments
$M_{2n}/M_{2n-2}$ calculated from \eref{moment ratio} for $r_w=1/2$
and $\sigma=1$.}
\label{Fig5}
\end{figure}

Computing the moments assuming \eref{LD tail} for all $v$, one gets
$\langle v^{2n} \rangle = (\sigma^2/2)^n (1-r_w^2)^{-n}
(2n)!/n!$. Certainly this results cannot be valid for small $n$ ,
e.g., compare the $n=1$ case with $2\mathrm{e}$ form \eref{energy}.
However, for large $n$ one expects this result to agree with the exact
result. To compare with the exact result, it is useful to look at the
ratio between two successive even moments, $\langle
v^{2n} \rangle/ \langle v^{2n-2} \rangle \sim 2\sigma^2 (1-r_w^2)^{-1}
n $ for large $n$. We find that the even moments $M_{2n}= \langle
v^{2n} \rangle$ satisfies the exact recursion relation
\begin{align}
\label{moment ratio}
&\Bigl[1-\epsilon^{2n} - (1-\epsilon)^{2n} +\gamma \bigl(1-r_w^{2n}\bigr)\Bigr]\,
M_{2n}=
\notag \\
&\sum_{m=1}^{n-1}\binom{2n}{2m}\epsilon^{2m}(1-\epsilon)^{2n-2m}M_{2m}M_{2n-2m}
\notag
\\
&+\gamma\sum_{m=0}^{n-1}\binom{2n}{2m} r_w^{2m} M_{2m}
\frac{(2n-2m)!}{(n-m)!} \left(\frac{\sigma^2}{2}\right)^{n-m},
\end{align}
with $M_0=1$. Using this we compute the moments
recursively. \Fref{Fig5} confirms that $M_{2n}/M_{2n-2}\rightarrow
2\sigma^2 (1-r_w^2)^{-1} n $ for large $n$.

Interestingly, \eref{recursion} has an exact solution, when the
external noise is drawn from the Cauchy distribution, i.e.,
$f(ik)=\exp(-a|k|)$. In this case, it can be proved that
$Z_c(k)=Z(ik)=\exp(-b|k|)$ with $b=a/(1-r_w)$, which  implies the Cauchy
law
\begin{equation}
P(v)=
\frac{1}{\pi b[1+(v/b)^2]}.
\end{equation}
Note that this result is independent of the rate parameter $\gamma$
and the coefficient of restitution of the inter-particle collisions.
It only depends on the coefficient of restitution $r_w$ for the
wall-particle collision and the noise parameter $a$.

In conclusion, in this letter we have shown that it is possible to
write exact recursion relation for the time evolution of the second
moment and velocity correlation together for the driven inelastic
Maxwell model. This enables us to obtain the form of the high-energy
tails and all moments (recursively) of the velocity distribution in
the steady state in the thermodynamic limit. We emphasize that we do
not require any approximations to break the BBGKY hierarchy, as has
been assumed in earlier work \cite{Costantini:07} but is unnecessary
--- the equations close once we also consider
correlations~\cite{Prasad}.  Our results are also valid in the
continuum time dynamics.

\acknowledgments
S.S. thanks S. N. Majumdar for useful discussions and acknowledges the
support of the Indo-French Centre for the Promotion of Advanced
Research (IFCPAR/CEFIPRA) under Project No. 4604-3.

\end{document}